# Polarization-insensitive space-selective etching in fused silica induced by picosecond laser irradiation

Xiaolong Li,[1,2] Jian Xu,[1,2,5] Zijie Lin,[1,2] Jia Qi,[3] Peng Wang,[3] Wei Chu,[3] Zhiwei Fang,[1,2] Zhenhua Wang,[1,2] Zhifang Chai,[1,2] and Ya Cheng[1,2,4,6]

[1]*State Key Laboratory of Precision Spectroscopy, School of Physics and Materials Science, East China Normal University, Shanghai 200062, China*
[2]*XXL—The Extreme Optoelectromechanics Laboratory, School of Physics and Materials Science, East China Normal University, Shanghai 200241, China*
[3]*State Key Laboratory of High Field Laser Physics, Shanghai Institute of Optics and Fine Mechanics, Chinese Academy of Sciences, Shanghai 201800, China*
[4]*Collaborative Innovation Center of Extreme Optics, Shanxi University, Taiyuan, Shanxi 030006, China*
[5]*jxu@phy.ecnu.edu.cn*
[6]*ya.cheng@siom.ac.cn*

**Abstract:** It is well known that when the fused silica is irradiated with focused femtosecond laser beams, space selective chemical etching can be achieved. The etching rate depends sensitively on the polarization of the laser. Surprisingly, we observe that by chirping the Fourier-transform-limited femtosecond laser pulses to picosecond pulses, the polarization dependence of the etching rate disappears, whereas an efficient etching rate can still be maintained. Observation with a scanning electron microscope reveals that the chirped pulses can induce interconnected nanocracks in the irradiated areas which facilitates efficient introduction of the etchant into the microchannel. The reported technology is of great use for fabrication of three-dimensional (3D) microfluidic systems and glass-based 3D printing.

## 1. Introduction

Femtosecond (fs) laser has become a powerful tool for fabricating three-dimensional (3D) microstructures in transparent materials, which is enabled in either additive/subtractive manners or an internal modification fashion [1-3]. For glass materials, the subtractive fabrication can be realized through irradiating the glass with focused fs laser beams to induce a space selective etching in chemical etchants [4-8]. This approach has been used for fabricating various microfluidic devices and optical components [9-11]. Intensive investigations reveal that for fused silica, the highest selectivity in the etching rate results from formation of self-organized nanograting structures oriented perpendicular to the polarization direction of the fs laser beams, which facilitates to efficiently introduce the etchants into the fabricated microchannels [12-16]. The polarization dependence of the etching selectivity requires the use of circular polarization or adaptive rotation of the polarization direction in writing 3D microfluidic channels, giving rise to additional complexity in the fabrication process [15, 16].

Here we report on an interesting and extremely useful observation which appears against the consensus of highly selective etching facilitated by nanograting formation as mentioned above. We systematically examine the dependence of etching rate on the laser polarization at different pulse durations in the range between hundreds of femtoseconds and ten picoseconds. Our results reveal an unexplored regime of interaction of intense ultrashort laser pulses with glass where neither nanogratings nor nanovoids but randomly oriented nanocracks are formed. More importantly, we achieve a high selectivity in the etching rate which is insensitive to the laser polarization. The observed characteristic is useful for producing various kinds of microfluidic structures of complex 3D geometries in glass.

## 2. Experimental

In our experiments, fused silica glass samples (Corning 7980) with a size of 10 mm×10 mm×2 mm were used as processing substrates. As illustrated in Figure 1a, an ultrashort laser system (Light Conversion, Pharos-20W) with a central wavelength of 1026 nm, a repetition rate of 50 kHz, a variable pulse duration ranging from 270 fs to 10 ps was employed for laser direct writing. The laser beam was focused 300 μm underneath the glass surface with an objective lens possessing a numerical aperture (NA) of 0.45. The average power of laser beam was tuned using a variable neutral density filter. To evaluate the dependence of the etching rate on various combinations of pulse duration, polarization and laser power, we scanned groups of lines (each group contains five parallel lines) with a length of 10 mm and a line spacing of 50 μm under the different conditions. The five lines in each group were written under the same condition, and the measurements were performed by averaging the data obtained from the individual lines in the same group. The pulse durations of the laser beam were set at 0.27, 1, 2, 4, 6, 8, 10 ps when the laser powers were set at 100, 200, 300, 400 mW, respectively. To evaluate the influence of linear polarization on the subsequent chemical etching, the θ is defined as the angle between the direct writing direction and the polarization direction. Laser writing in both polarization directions (⊥, θ=90°) and (∥, θ=0°) were performed (see Figure 1b). Moreover, a quarter waveplate was inserted in the light path shown in Figure 1a when the circularly polarized beam (O, circular) was used. The writing speed was maintained at 500 μm/s throughout the experiment.

After the laser irradiation, the glass samples were grinded and polished to remove the end of the laser-written tracks as the focal spot of the laser beam was spoiled when approaching the end facets due to refractive index mismatch at glass/air interfaces. Then, the samples were immersed in a 10 mol/L KOH solution to undergo an ultrasonic bath at 85 °C for 1 hour. After the chemical etching, the glass samples were examined by an optical microscopy (Olympus, BX53), by which the length of the etched channels were determined. To investigate the modification mechanisms under the different irradiation conditions, the modified glass samples were grinded and polished to expose the modification tracks. The morphology of the modification tracks was revealed by briefly etching the polished samples in a 10 mol/L KOH solution for 10 min. Afterwards, the nanostructures formed in the laser irradiated regions were observed by SEM (Hitachi, S4800).

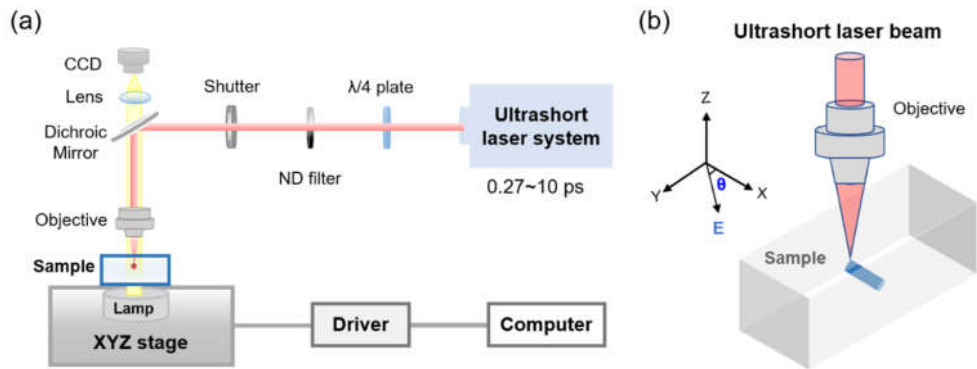

Fig. 1. Schematic of (a) the experimental setup for ultrashort laser processing and (b) laser direct writing in fused silica for modification.

## 3. Results

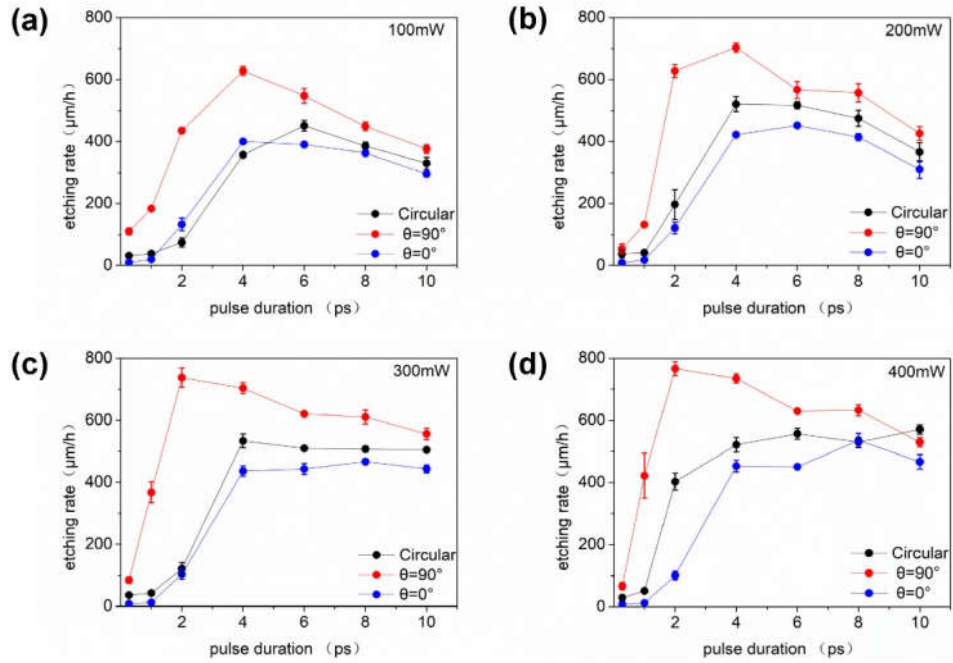

Fig. 2. Etching rates of laser modified lines in fused silica versus pulse durations at different polarization conditions and laser powers: (a) 100 mW; (b) 200 mW; (c) 300 mW; (d) 400 mW.

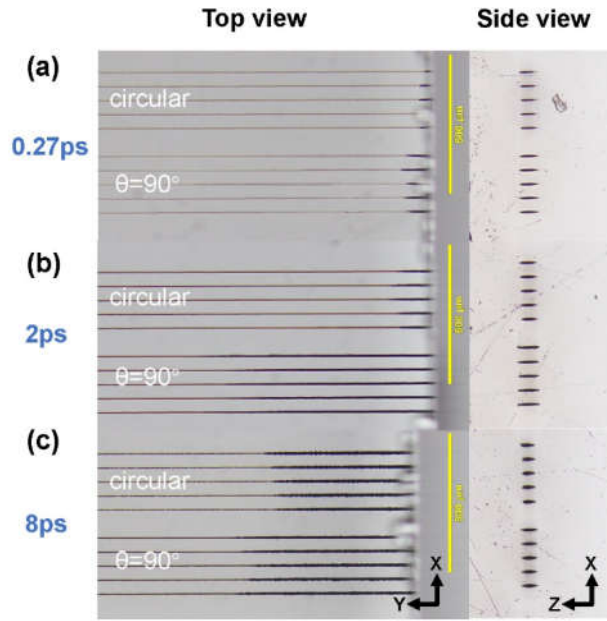

Fig. 3. Top-view and side-view optical micrographs of etched channels at different pulse durations and polarization conditions for 1 h, 300 mW. (a) 0.27 ps; (b) 2ps; (c) 8 ps.

Figure 2 presents the 1-hour KOH etching efficiencies of the laser modified lines under the different irradiation conditions. The influence of pulse durations on the etching rates can be divided into two regimes: 0.27 ps (270fs) ~ 4ps and 4~10 ps. In the regime between 0.27 ps and 4 ps, one can see a strong dependence of the etching rate on the polarization of the laser beam. The etching rates obtained with linearly polarized ($R_\perp$, $\theta=90°$) laser beams are significantly higher than that obtained with the circularly polarized laser beams ($R_O$, circular) and the linearly polarized ($R_{//}$, $\theta=0°$) laser beams. In particular, an etching rate of 766.3±22.2 μm/hour can be obtained with the linearly polarized laser beams ($R_\perp$) at 2ps when the laser power is set at 400 mW (see Figure 1d and Table 1), which is the highest etching rate obtained in our investigation. It is well known that the dependence of etching rate on the writing laser polarization originates from the formation of nanogratings in fused silica under the repeated irradiation with ultrashort laser pulses [17-20]. It can be inferred that the etching behaviors in this regime is dominated by the existence of nanogratings induced by ultrashort laser pulses.

Table 1. Etching rates (R) with different and pulse durations at 400 mW and the ratios of the etching rates under the different polarization conditions ($\perp$, //, O)

| Pulse duration | 0.27ps | 1ps | 2ps | 4ps | 6ps | 8ps | 10ps |
|---|---|---|---|---|---|---|---|
| $R_\perp$, $\theta=90°$ | 67.1±12.0 | 422.2±72.0 | 766.3±22.2 | 734.4±14.5 | 629.9±6.7 | 633.1±17.7 | 530.0±14.8 |
| $R_{//}$, $\theta=0°$ | 8.7±1.5 | 12.9±1.0 | 101.1±14.0 | 453.2±18.5 | 450.7±3.6 | 535.0±23.2 | 466.9±23.5 |
| $R_o$, circular | 29.9±9.5 | 52.0±6.4 | 402.5±27.7 | 521.4±23.3 | 556.8±17.7 | 529.8±15.4 | 570.6±15.2 |
| $R_\perp$: $R_{//}$: $R_o$ | ~7.7:1:3.4 | ~32.7:1:4 | ~7.6:1:4 | ~1.6:1:1.2 | ~1.4:1:1.2 | ~1.2:1:1 | ~1.1:1:1.2 |

In contrast, in the regime between 4 ps and 10 ps of the pulse durations, the etching results appear very different. The etching rates obtained with the circularly polarized laser beams ($R_O$) and the linearly polarized ($R_{//}$) laser beams dramatically increase as shown in Figure 2, whereas the rate obtained with the linearly polarized ($R_\perp$) laser beams decreases. Eventually when the pulse duration reached 10 ps, the dependence of etching rate on the writing laser polarization almost disappear (i.e., without noticeable difference in the two cases). The optical microscope images of 1-hour KOH etching of modified structures at 8ps, 300 mW shown in Figure 3c clearly demonstrates a polarization-insensitive etching as compared with the results in Figures 3a (0.27 ps) and 3b (2ps). As shown in Figure 1d and Table 1, when the pulse duration is increased to 8 ps at 400 mW, the average etching rates achieved with the linearly polarized laser beams $R_\perp$ and $R_{//}$ as well as the circularly polarized laser beam $R_O$ are ~633, ~535, and ~530 μm/hour, respectively. The ratio of the corresponded etching rates obtained under the different polarization conditions ($R_\perp$: $R_{//}$: $R_O$) with the 8 ps pulses is only ~1.2:1:1, which is very different from the ratio of ~7.6:1:4 obtained with the 2 ps pulses (see Table 1). The disappearance of polarization sensitivity is an indication of the disappearance of the nanogratings which will be revealed below. Moreover, when the laser power increases from 200 mW to 400 mW in the regime between 4 ps and 10 ps, the etching rates tend to be saturated, which is also different from the case of the regime between 0.27 ps and 4 ps (Figures 2b, 2c and 2d). Therefore, the etching behavior in the regime between 4 ps and 10 ps is dominated by a new mechanism instead of the nanograting formation.

## 4. Discussion

To understand the high etching rates independent of the laser polarization obtained in fused silica with the ps laser irradiation, we polished the irradiated samples to expose the laser modified regions. The samples were then undergone a brief chemical wet etching in KOH for 10 min to disclose the nanometer scale morphologies in the modified structures which are shown in Fig. 4. It can be seen that when irradiated with the linearly polarized fs laser pulses, nanogratings are formed in fused silica which help accelerate the refreshment of etchant in the fabricated microchannels (see the left and middle panels of Fig. 4a). The well-organized nanogratings are transformed to random nanostructures when the linear polarization of the fs laser pulses is switched to circular polarization, as shown in the right panel of the Fig.4a. These results are consistent with the previous reports on the fs laser induced selective etching in fused silica. Moreover, when the fs laser pulses are extended to 1 ps and 2 ps as shown in Figs. 4b and 4c, instead of the familiar nanogratings randomly oriented nanocracks start to appear in the laser modified area. Furthermore, when the pulse duration is extended to 4 ps, the random nanocracks prevail against the nanograting structures (see Fig. 4d). Therefore, the highest etching rate achieved at 2 ps pulse duration could be ascribed to the synergetic contribution from the nanogratings and the interconnected random nanocracks. Moreover, the determinative role of nanograting formation in the ultrashort laser induced selective etching in the regime between 0.27 ps and 4 ps is confirmed.

In contrast, when the laser pulse duration increases to above 4 ps, the laser modified areas are dominated with the randomly oriented nanocracks instead of the nanogratings, as shown in Figs. 4e and 4f. It is reported that the nanograting structures could still be formed even with 8 ps pulses at a higher repetition rate (200 kHz) when a higher NA (0.65) objective lens was used [20]. Thus, it is worth noting that the results presented here are dependent of the focal condition and laser repetition rate. The physical mechanism behind the formation of such nanocracks is unclear which might be associated with the different energy deposition as well as the stress buildup mechanisms in the interaction of picosecond laser with the transparent dielectrics as compared with the fs laser interaction counterpart [21-27]. The high stress is also indicated by the fact that the nanocracks in the laser modified areas are preferentially oriented along the laser scan direction when irradiated with the circularly polarized pulses (Figs. 5d, 5e and 5f). Nevertheless, from an application of point of view, the new interaction regime enabled with the picosecond laser pulses for polarization-insensitive space-selective etching is of great use for fabrication of 3D microstructures in fused silica, as the necessity of dynamic adjustment of direction of the laser polarization is eliminated (see Table 1). To showcase such a capability, we fabricated an array of triangle loops of microchannels in fused silica using linearly and circularly polarized fs and picosecond laser pulses and compared the results as shown in Fig. 5. Clearly, the picosecond laser direct writing in the regime of 4~10 ps (Figs. 5c, 5d, 5e and 5f) provides uniform etching rates in all the segments regardless of their orientations, whereas in the regime of 0.27~2 ps (Figs. 5a and 5b), only the segment orientated in the same direction as that of the nanograting can be etched at a decent removal rate, which are consistent with the results in Figure 2.

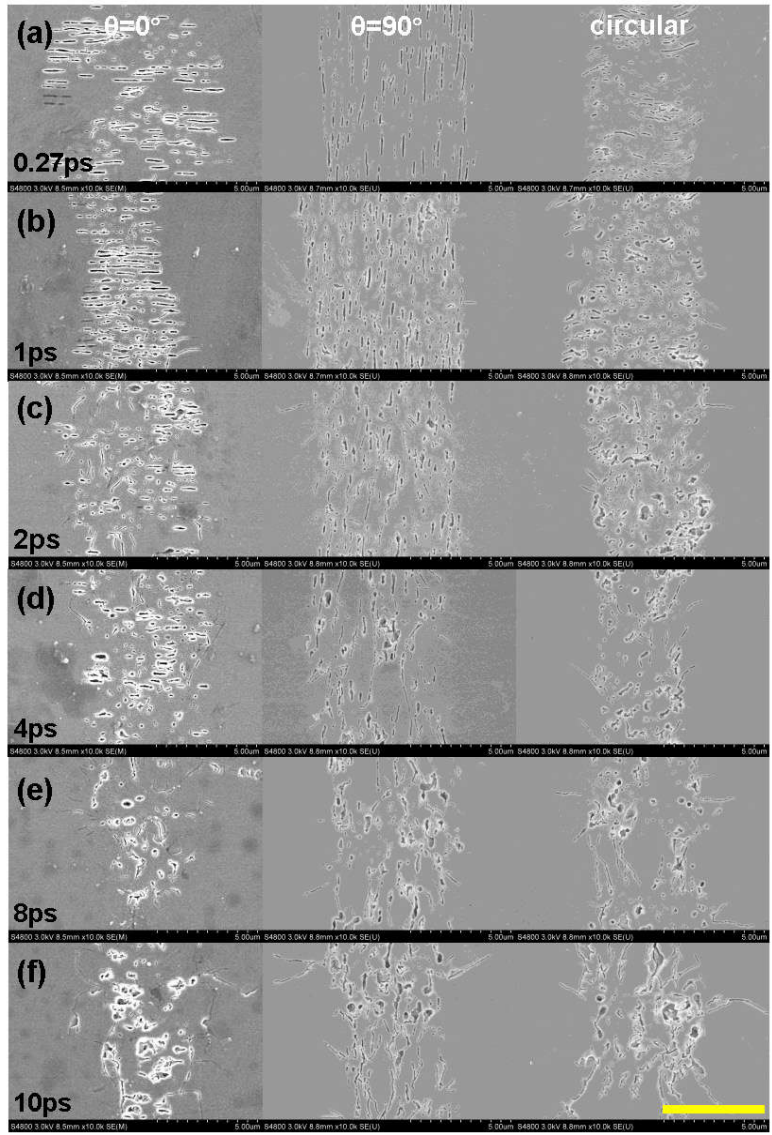

Fig. 4. SEM images of the irradiated area at different pulse duration and different polarization conditions at 400 mW. (a) 0.27 ps; (b) 1 ps; (c) 2ps (d) 4 ps (e) 8 ps (f) 10 ps. Polarization from left to right in each figure: ⊥, ∥, O. Scale bar is 5 μm.

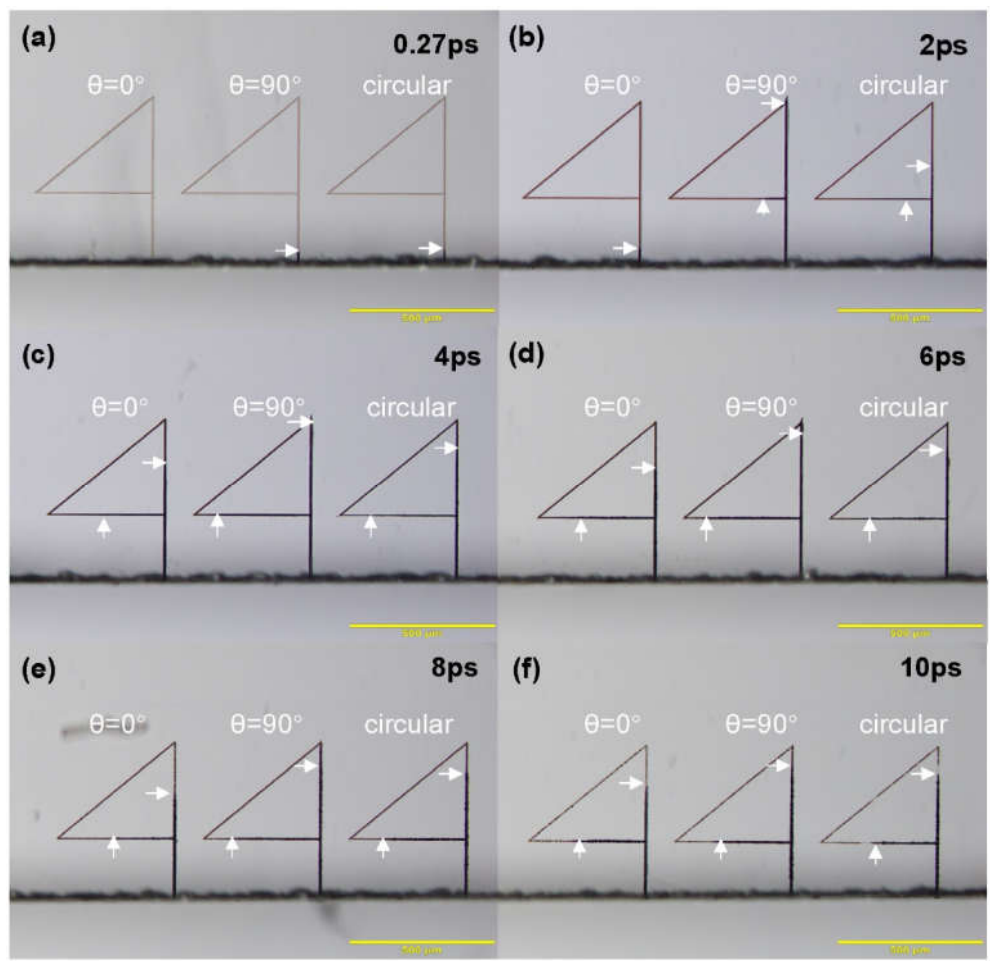

Fig. 5. The etching of a loop-channel of triangle shape undergone the different irradiation condition at 400 mW (a) 0.27 ps; (b) 2 ps; (c) 4ps (d) 6 ps (e) 8 ps (f) 10 ps.

## 5. Conclusion

To conclude, our investigations reveal a new region of interaction of ultrashort laser pulses with fused silica which does not form either smooth modification (i. e., region 1) and nanograting (i.e., region 2) or nanovoid (i.e., region 3) as reported previously [1,10,17-20]. It is shown that the picosecond laser irradiation can produce interconnected nanocracks which are preferentially oriented along the laser scan direction. The revealed characteristic is used for fabricating microfluidic channels along different directions to achieve the high etching rates insensitive to the writing laser polarization state.

**Funding**

National Natural Science Foundation of China (Grants 61761136006, 11734009, 61590934, 11674340, and 61505231), Strategic Priority Research Program of the Chinese Academy of Sciences (Grant XDB16030300), Key Research Program of Frontier Sciences, Chinese Academy of Sciences (Grant QYZDJ-SSW-SLH010), Project of Shanghai Committee of Science and Technology (Grant 17JC1400400), and Shanghai Pujiang Program (Grant 18PJ1403300).

**References:**
1. K. Itoh, W. Watanabe, S. Nolte, and C. B. Schaffer, "Ultrafast processes for bulk modification of transparent materials," MRS Bull. 31(08), 620–625 (2006).
2. R. R. Gattass and E. Mazur, "Femtosecond laser micromachining in transparent materials," Nat. Photon. 2(4), 219–225 (2008).
3. K. Sugioka and Y. Cheng, "Ultrafast lasers-reliable tools for advanced materials processing," Light: Sci. & Appl. 3(4), e149 (2014).
4. A. Marcinkevičius, S. Juodkazis, M. Watanabe, M. Miwa, S. Matsuo, H. Misawa, and J. Nishii, "Femtosecond laser-assisted three-dimensional microfabrication in silica," Opt. Lett. 26(5), 277–279 (2001).
5. Y. Bellouard, A. Said, M. Dugan, and P. Bado, "Fabrication of high-aspect ratio, micro-fluidic channels and tunnels using femtosecond laser pulses and chemical etching," Opt. Express 12(10), 2120–2129 (2004).
6. C. Hnatovsky, R. S. Taylor, E. Simova, P. P. Rajeev, D. M. Rayner, V. R. Bhardwaj, and P. B. Corkum, "Fabrication of microchannels in glass using focused femtosecond laser radiation and selective chemical etching," Appl. Phys. A 84(1-2), 47–61 (2006).
7. S. Kiyama, S. Matsuo, S. Hashimoto, and Y. Morihira, "Examination of etching agent and etching mechanism on femtosecond laser microfabrication of channels inside vitreous silica substrates," J. Phys. Chem. C 113(27), 11560–11566 (2009).
8. J. Gottmann, M. Hermans, N. Repiev, and J. Ortmann, "Selective laser-induced retching of 3D precision quartz glass components for microfluidic applications-up-scaling of complexity and speed," Micromachines 8(4), 110 (2017).
9. K. Sugioka, J. Xu, D. Wu, Y. Hanada, Z. Wang, Y. Cheng, and K. Midorikawa, "Femtosecond laser 3D micromachining: a powerful tool for the fabrication of microfluidic, optofluidic, and electrofluidic devices based on glass," Lab Chip 14(18), 3447–3458 (2014).
10. D. Choudhury, W. T. Ramsay, R. Kiss, N. A. Willoughby, L. Paterson, A. K. Kar, J. Nishii, and G. Cerullo, "A 3D mammalian cell separator biochip," Lab Chip 12(5), 948–953 (2012).
11. M. Haque, K. K. C. Lee, S. Ho, L. A. Fernandes, P. R. Herman, Y. Shimotsuma, K. Miura, K. Hirao, P. V. S. Marques, D. Kopf, G. Mayer, J. Albert, M. Rothhardt, C. Krafft, and J. Popp, "Chemical-assisted femtosecond laser writing of lab-in-fibers," Lab Chip 14(19), 3817–3829 (2014).
12. C. Hnatovsky, R. S. Taylor, E. Simova, V. R. Bhardwaj, D. M. Rayner, and P. B. Corkum, "Polarization-selective etching in femtosecond laser-assisted microfluidic channel fabrication in fused silica," Opt. Lett. 30(14), 1867–1869 (2005).
13. M. Beresna, M. Gecevičius, and P. G. Kazansky, "Polarization sensitive elements fabricated by femtosecond laser nanostructuring of glass," Opt. Mater. Express 1(4), 783–795 (2011).
14. M. Hermans, J. Gottmann, and F. Riedel, "Selective, laser-induced etching of fused silica at high scan-speeds using KOH," J. Laser Micro Nanoeng. 9(2), 126–131 (2014).
15. X. M. Yu, Y. Liao, F. He, B. Zeng, Y. Cheng, Z. Z. Xu, K. Sugioka, and K. Midorikawa, "Tuning etch selectivity of fused silica irradiated by femtosecond laser pulses by controlling polarization of the writing pulses," J. Appl. Phys. 109(5), 053114 (2011).
16. C. A. Ross, D. G. Maclachlan, D. Choudhury, and R. R. Thomson, "Optimisation of ultrafast laser assisted etching in fused silica," Opt. Express 26(19), 24343–24356 (2018).
17. Y. Shimotsuma, P. G. Kazansky, J. Qiu, and K. Hirao, "Self-organized nanogratings in glass irradiated by ultrashort light pulses," Phys. Rev. Lett. 91(24), 247405 (2003).
18. R. Taylor, C. Hnatovsky, and E. Simova, "Applications of femtosecond laser induced self-organized planar nanocracks inside fused silica glass," Laser Photonics Rev. 2(1-2), 26–46 (2008).
19. S. Richter, M. Heinrich, S. Doring, A. Tunnermann, S. Nolte, and U. Peschel, "Nanogratings in fused silica: Formation, control, and applications," J. Laser Appl. 24(4), 042008 (2012).
20. C. Corbari, A. Champion, M. Gecevičius, M. Beresna, Y. Bellouard, and P. G. Kazansky, "Femtosecond versus picosecond laser machining of nano-gratings and micro-channels in silica glass," Opt. Express 21(4), 3946–3958 (2013).
21. H. Zhang, S. M. Eaton, and P. R. Herman, "Low-loss Type II waveguide writing in fused silica with single picosecond laser pulses," Opt. Express 14(11), 4826–4834 (2006)
22. B. McMillen, B. T. Zhang, K. P. Chen, A. Benayas, and D. Jaque, "Ultrafast laser fabrication of low-loss waveguides in chalcogenide glass with 0.65 dB/cm loss," Opt. Lett. 37(9), 1418–1420 (2012).
23. L. Capuano, R. Pohl, R. M. Tiggelaar, J. W. Berenschot, J. G. E. Gardeniers, and G. R. B. E. Römer, "Morphology of single picosecond pulse subsurface laser-induced modifications of sapphire and subsequent selective etching," Opt. Express 26(22) 29283–29295 (2018).
24. D. Nieto, J. Arines, G. M. O'Connor, and M. T. Flores-Arias, "Single-pulse laser ablation threshold of borosilicate, fused silica, sapphire, and soda-lime glass for pulse widths of 500 fs, 10 ps, 20 ns," Appl. Opt. 54(29), 8596-8601 (2015).
25. S. Ly, N. Shen, R. A. Negres, C. W. Carr, D. A. Alessi, J. D. Bude, A. Rigatti, and T. A. Laurence, "The role of defects in laser-induced modifications of silica coatings and fused silica using picosecond pulses at 1053 nm: I. Damage morphology," Opt. Express 25(13), 15161–15178 (2017).

26. T. A. Laurence, R. A. Negres, S. Ly, N. Shen, C. W. Carr, D. A. Alessi, A. Rigatti, and J. D. Bude, "Role of defects in laser-induced modifications of silica coatings and fused silica using picosecond pulses at 1053 nm: II. Scaling laws and the density of precursors," Opt. Express 25(13), 15381–15401 (2017).
27. K. Bergner, B. Seyfarth, K. A. Lammers, T. Ullsperger, S. Döring, M. Heinrich, M. Kumkar, D. Flamm, A. Tünnermann, and S. Nolte, "Spatio-temporal analysis of glass volume processing using ultrashort laser pulses," Appl. Opt. 57(16), 4618–4632 (2018).